\documentclass[preprint2]{aastex}

\shorttitle{Extragalactic Jets}
\shortauthors{De Young}

\begin{document}


\title{The Particle Content of Extragalactic Jets}


\author{David S. De Young}
\affil{NOAO, Tucson, AZ 85719}


\begin{abstract}
Recent radio and x-ray data from radio sources in galaxy clusters are used
to place constraints on the particle content and kinematics of jets that
supply these radio sources.  These data show that the $pdV$ work required
to inflate the radio lobes in the ICM exceeds all other estimates of the
amount of energy needed to supply the observed radio emission.  If
the required jet energy density has an isotropic
pressure,
then in almost all cases the
jets cannot be confined by the external pressure of the ICM.  This problem
can be resolved with jets dominated
by a cold and relatively dense proton population, but even here the
accompanying energy density in electrons alone can cause jet decollimation
in many cases.  Calculation of particle interactions in a 
cold proton jet shows the electron-proton energy transfer times to be
very long.
Electron-positron jets, unless highly beamed and
with an unusual energy distribution, cannot solve the decollimation problem.
A viable alternative may be "Poynting Flux" dominated
jets with a very low particle content.
\end{abstract}


\keywords{radio galaxies, jets, galaxy clusters}


\section{Introduction and Motivation}

Although collimated bipolar outflows from active galactic nuclei have been
studied for over forty years, many fundamental characteristics of these
outflows are still poorly known or completely unknown. Among these are the
outflow speed, the energy flux of the outflow, and the nature and composition
of the outflowing material (e.g., De Young 2002).  In  
particular, the particle content and the total energy flux of the outflows,
both key parameters to characterizing the nature of the "central engine"
that creates them, are poorly known.  Radio observations require
only the presence of a relativistic electron population and an
accompanying magnetic field to produce the observed synchrotron radiation.
There is presumably another charged
particle species carried outward with the electrons and field that
provides charge neutrality, but its nature has remained generally
obscure and observationally inaccessible.
Hence indirect arguments have been put forward to support various
possibilities, such as positrons, protons, and possibly other more
exotic species (e.g., Ghisellini et al. 1993;
Celotti \& Fabian 1993; Reynolds et al. 1996).
Because the energy flux carried by this positively 
charged population has had no clear 
empirical constraints, its value has been usually set by assumption
or by past practice. 

However,
during the last few years a combination of x-ray and radio data has emerged
that provides for the first time some unambiguous measurements of the
characteristics of extragalactic jets.
These observations are of extended radio sources that reside in rich
clusters of galaxies, and the data reveal the presence of cavities in
the intracluster medium (ICM) that are coincident with the extended lobes of
radio emission associated with an active galaxy within the cluster. 
(e.g., B\"ohringer et al. 1993; Fabian et al. 2000, 2003a, 2003b;
McNamara et al. 2000, 2001; David et al. 2001; Nulsen et al. 2002;
Birzan et al. 2004; Blanton et al. 2004; Kempner et al. 2004).
The inference drawn from these data is that the cavities in the ICM have been
inflated by the expanding radio lobes; in some cases the extended radio
emitting regions show evidence of ongoing injection of energy from jets
emanating from the nucleus, while in other cases the cavities are associated
with relic radio lobes.  In many cases there is evidence suggesting that
the lobes are currently expanding very slowly, if at all, into the ICM,
and that the radio "bubbles" may be rising buoyantly through the 
intracluster medium. 
Many ideas have
been put forward in interpreting the data or in modeling the dynamics
of these objects (e.g., Quilis et al. 2001; Brighenti \&
Mathews 2002; Br\"uggen 2003; De Young 2003). 
Hydrodynamic simulations of
the rise of buoyant bubbles have been carried out in two and three
dimensions (e.g., Churazov et al. 2001; Br\"uggen et al. 2002;
Br\"uggen \& Kaiser 2002; Reynolds et al. 2002), 
and more realistic MHD simulations have also been performed
(Br\"uggen \& Kaiser 2001; Robinson et al. 2004; Jones \& De Young 2005). 

An essential aspect of these observations relevant 
to the content of extragalactic jets 
is that the $pdV$ work required to form 
the cavity in the ICM can be directly measured, since the density
and temperature of the ICM are known.  Hence for the first time an
unambiguous measure of the minimum total energy injected into the
expanding radio sources can be made, and the consequences of this
remarkable calorimetry in understanding the nature of the radio sources
themselves is the subject of this paper.  Section 2 derives some 
constraints that can be placed on the 
energetic and dynamics of jets which come directly from
the $pdV$ values derived from observations.
Section 3 considers some consequences of the
results of Section 2 in terms of the content of outflowing jets, 
and Section 4 discusses particle interactions within jets. 
Section 5 provides some additional
overall conclusions and consequences of this work.

\section{Observational Constraints}

In this section observational data are used in
conjunction with a minimal set of additional assumptions to determine
constraints that can be placed on the content of 
extragalactic jets.  The x-ray 
data and the radio morphology data permit a direct determination of
the $pdV$ work needed to inflate the 
cavity in the ICM, with no additional assumptions
(e.g., McNamara et al. 2001; Birzan et al. 2004).
The radio flux, source distance 
and spectral index provide a measure of the radio luminosity.
With just these numbers some interesting comparisons
can be made.  For example, if the value of $pdV$ is a measure of the
energy injected to date into the ICM by the active nucleus, 
how does this compare with the energy needed to power the radio source
at its current luminosity over some radiative or dynamical lifetime?
Is there enough $pdV$ energy to do this?  Is there an excess of
energy, and if so, what form has it taken? 
(The total energy injected is of course more than the value given by
$pdV$, since there is some internal energy present in the material
that has inflated the bubble.) 
How do these numbers compare to the
usual "equipartition" energies for the radio source?  If more energy
is needed to perform the $pdV$ work than is required to account for 
the radio emission, how is this energy transported from
the nuclear regions into the radio lobes?  Does consideration of this
energy transport place any constraints on the form the energy takes;
i.e., upon the content of the extragalactic jets?  It is this last
question that is the focus here.

\subsection{Total Energies}
The first step in providing some answers to these questions is to 
estimate the total energy $E_{e}$ 
needed in relativistic electrons to produce the radio
emission that is seen at the present time.  At least this much energy
must be produced in the nucleus and carried outward by the jets, and it
can then be compared to the $pdV$ energy measured by completely different
means. For an observed power law radio flux distribution of the form
$S(\nu) \propto \nu^{-\alpha}$, with $\alpha$ positive, the underlying
relativistic electron energy distribution 
per unit volume is of the form $n(E,\mathbf{r})dE = 
K E^{-p}dE$, where $\alpha = (p - 1)/2$.  Here the electron
distribution has been assumed homogeneous and isotropic, so that
$K$ is not a function of position $\mathbf{r}$ within the emitting 
region.  
The total energy in electrons that is needed is then just
$E_{e} = \int n(E,\mathbf{r}) E dE d\mathbf{r} = K V \int E^{-p+1}dE$,
and the problem is then to determine the value of the coefficient $K$
in the electron energy distribution.  This is done by first writing 
the intensity of synchrotron radiation $I_{\nu}$
emitted along a line of sight.  This expression, which contains $K$,
is then integrated over the emitting volume to obtain a value for
the total radio flux $S_{\nu}$.  For a
power law flux distribution of the above form, $S_{\nu}$ is 
related to an observed flux $S_{o}$ at some frequency $\nu_{o}$
by $S_{\nu} = S_{o}(\nu/\nu_{o})^{-\alpha}$, and this expression is
then solved for $K$ and the result inserted into the equation for
$E_{e}$.
The derivations of $I_{\nu}$, $S_{\nu}$, and $E_{e}$ are well known 
(e.g., Ginzburg \& Syrovatski 1964; De Young 2002),
but care needs to be exercised in 
integrating over the frequency spectrum of the individual
electrons and in integrating over the pitch angle distribution of the
electrons (e.g., Korchak 1957; Trubnikov 1958;  
Syrovatskii 1959).
Combining the derivations in these references gives, in CGS units,
\begin{eqnarray}
E_{e} = \frac{1.48 \times 10^{12}}{a(p)}
D^{2} B^{-3/2} \times \nonumber \\
  \frac{S_{o}\nu_{o}^{\alpha}}
{1/2 - \alpha}(\nu_{U}^{(1/2-\alpha)}
- \nu_{L}^{(1/2 - \alpha)}),
\end{eqnarray}
where $D$ is the distance to the source and $\nu_{L}$ and 
$\nu_{U}$ are the lower and upper frequency limits.  The 
function $a(p)$ contains the coefficients resulting from
integration over electron pitch angles and over the emitted
radiation of individual electrons. The values of $a(p)$ range
from 0.28 to 0.07 as $p$ varies from 1 to 4, so this variation
of $a(p)$ can have meaningful consequences. 
A more approximate treatment that does not integrate over 
the pitch angle distribution or the emission spectrum of 
individual electrons eliminates $a(p)$, 
and the coefficient becomes a constant  
approximately equal to $4\pi \times 10^{12}$ in the same cgs units 
(Lang 1980, Fabian et al. 2002).

In order to proceed further the magnetic field in Eq. 1 must be 
specified, and there are no independent observational constraints
available to determine the mean value of $B$.  A common
practice is
 to assume rough energy equipartition between the
magnetic and particle energies in a radio lobe, 
since this minimizes the total energy required.
This total lobe energy is
$E_{tot} = E_{p} + E_{B}$, where $E_{p}$ is the energy of all
particles present (the "plasma"), and $E_{B}$ is the total 
magnetic field energy.  The energy in electrons (Eq. 1) can be
written as $f B^{-3/2}$, and it is usually assumed that
the additional energy in the charge neutralizing species (and any
others) can be written as some multiple $\kappa$ of $E_{e}$, so
that $E_{p} = (1+\kappa)E_{e}$.  For a field that is homogeneous
and isotropic on scales comparable to the source size, 
$E_{B} = V(B^{2}/8\pi)$, where $V$ is the source volume and 
$B$ is some appropriate average value of the magnetic field.  In addition
a filling factor $\phi$ for the "relativistic plasma" in the lobe
can be introduced into $E_{B}$ (e.g., Fabian et al. 2002; 
Dunn \& Fabian 2004; Dunn et al. 2005),
but in view of the unknown nature of the interstitial material,
it is not clear that firm new constraints on the jet content will be
obtained from values of $\phi \ne 1$. 
Differentiation of $E_{tot}$ with respect to $B$ 
and finding the extremum value gives the
standard equipartition result:
\begin{equation}
B_{eq} = \left[ \frac{8 \pi (1 + \kappa) f}{V}\right]^{2/7}
\end{equation}

It seems unlikely that the magnetic energy will exceed this equipartition
value, since in that case the magnetic field will control the dynamics
of the radio source evolution, and models of magnetically dominated
jets have not yet been shown to be stable over long periods of time and
over distances of tens to hundreds of jet radii (e.g.,
Begelman 1998; Li 2002; Hsu \& Bellan 2002).
(This applies to MHD
models and does not include "Poynting Flux" models (e.g.
Lovelace et al. 2002).)
The field strengths 
could be well below the equipartition value, and this has been argued
to be the case for several cluster radio sources
(e.g., Fabian et al. 2002; Dunn \& Fabian 2004).
Of course, as the magnetic field strength decreases below equipartition,
the required electron energy rapidly rises, as can be seen in Eq. 1.
However, the electron energy cannot rise without limit, as will be
discussed in Section 2.4, and so this sets 
lower limits to the departure from
equipartition of $E_e$.  Another 
issue is equipartition with respect to the entire
particle population or with respect to the relativistic electrons only.
This is relevant if the charge neutralizing species
that is co-located with the electrons carries significantly more energy
per particle than do the electrons.  In this case the magnetic field
could be in near equipartition with the electrons and yet be well below
equipartition with the particle population in general.  If this is so,
then the magnetic field is given by $B_{eq}$ with $\kappa = 0$,
and this value can then be used in Eq. 1 to determine a value for the
total electron energy.

There has previously been no observational motivation for choosing a
value for the energy contained in the positively charged particle
species.  While it may be true that "a typical value used in the
literature is k = 100" (Fabian et al. 2002) 
this value arises from consideration of cosmic rays in our own Galaxy
(Burbidge 1956, 1959) 
and has no relation to extragalactic radio sources. If the positively
charged species is positrons, then it is likely that they will be
in energy equipartition with the relativistic electrons.
while if this species is protons, then there is no a priori value
that is more likely than any other.

\subsection{Timescales}
A second step is to estimate
the time during which the $pdV$ energy is injected into the ICM in
order to obtain the total energy fluxes required to be transported by
the collimated outflows.  This timescale is not known,
since the ages of the extended radio sources are not
known.  However, limits can be placed on the duration of the injection
process.  One limit is the electron radiative lifetime, unless
reacceleration is the {\it major} source of electron energy at
the present epoch.\footnote{Though there is indirect evidence
for reacceleration in the lobes of some large sources from spectral
index maps and particle transit time arguments (e.g. Mack et al. 1998),
the spectral index data are not unambiguous (Blundell \& Rawlings 2000;
Treichel et al. 2001) and the phenomenon is not generally seen in
small FR-I sources such as the ones in this sample.}
Absent significant reacceleration,
the outflow has occurred over a time less than the radiative
lifetime of the electrons, and a relativistic electron radiating at
a frequency $\nu$ loses one half its energy in a time given by
(e.g., De Young 2002)
\begin{equation}
t_{1/2} = \frac{9 m_{e}^{3}c^{5}}{4 e^{4}} 
\left( \frac{3 e}{4 \pi m_{e}c}\right)
^{1/2} B^{-3/2} \nu^{-1/2},
\end{equation}
or,
$t_{1/2} = 5.2 \times 10^{7} B_{-5}^{-3/2} {\nu_{9}^{-1/2}}$ yr.
Here $B_{-5}$ is the average
magnetic field strength in units of $10^{-5}$ G, and $\nu_{9}$ is
the frequency of the synchrotron radiation in GHz. 

A second time scale comes from estimates of dynamical processes involved
in these objects, and a time that has been frequently used in this
connection is the buoyant rise time of the source,
assuming sources form buoyant bubbles in the ICM.  Given
that assumption, buoyant rise times can be estimated 
from approximate buoyant speeds 
(e.g., Churazov et al. 2001),
where the time calculated is that for the radio emitting cavity to
rise buoyantly from the location of the active nucleus to its presently
observed position.  These lifetimes are also upper limits, for it is very
unlikely that the buoyant phase actually begins at the position of the
central energy source.  A more accurate 
buoyant lifetime would be the time required for the cavity to rise from
the termination radius of the outflowing jet (which is not known) to the
present position of the bubble.
The buoyant rise time obtained from the estimated buoyant velocity
given by Churazov et al. (2001) is 
$t_{buoy} = D/v_{b} \sim D/[2gV_{b}/{C_{D} A_{b}}]^{1/2}$,
where $D$ is the distance of the bubble from the parent nucleus,
$A_{b}$ is the bubble cross section, $V_{b}$ its volume, $g$ the
(average) gravitational acceleration, and $C_{D}$ is an estimated
drag coefficient.

\subsection{Energy Fluxes}

Given the total energy requirements from both the $pdV$ energy and the
total energy in electrons required to power 
the observed synchrotron radiation,
the above timescales allow an estimate of lower limits to the energy flux
carried by the jets.  Since the magnetic field dependence
of the total electron energy is the same as 
that for the radiative lifetime (Eqs. 1 and 3),
the energy flux required to produce the observed 
synchrotron emission during a radiative lifetime, $E_{e}/t_{rad}$, 
is independent of the
assumed value for the magnetic field. 
Although this estimate of the energy flux is free from assumptions of
magnetic field strength, a comparison 
needs to be made with the energy
flux needed to supply the $pdV$ work, which may not be
independent of $B$.  This latter energy flux 
can be found once an outflow time is chosen,
and logical choices are 
both the radiative and buoyant timescales.  
Use of a radiative timescale requires an
assumption about magnetic field strengths,
and the above discussion about
equipartition fields suggests that use of a field value that is in
equipartition with the electron energy density alone can provide an
interesting limiting case, given other arguments that suggest that the 
field is at or below overall equipartition (Fabian et al. 2002,
Dunn \& Fabian 2004).

Once these energy fluxes are in hand, parameters that
characterize the collimated outflow itself can be used to arrive at
the final set of constraints.  The two jet parameters that are
needed are the jet radius and the bulk flow speed of material within the jet.
These, when combined with the total energy flux required,
give the energy density $\epsilon_{j}$ within the jet, 
since the total energy flux is just 
$dE_{j}/dt = \epsilon_{j} v_{j} A_{j}$,
where $v_{j}$ is the
bulk flow speed in the jet, and $A_{j}$ is the jet cross section.
For radio sources and relics in clusters of galaxies 
the pressure in the ICM through which the
jet travels is known, and this external pressure can be
compared with the range of internal energy 
densities $\epsilon_{j}$ interior to the jet.
This comparison assumes 
that the internal energy density results in
an isotropic comoving pressure, such as that from a relativistic electron
population with random velocities.
Internal energy densities in excess of
the external values of $nkT$ mean that that particular jet cannot be 
confined by the ICM.  For these cases either the internal energy
density must be lowered by some means, or the internal energy
in the jet must not create an associated isotropic pressure,  
or an alternate confinement
mechanism must be invoked.  The most obvious candidate for 
alternate confinement is some
form of magnetic confinement, but as mentioned previously, 
these mechanisms require very special
initial conditions, very strong anisotropic magnetic fields, and they
are often unstable.  Since magnetic confinement has not yet been shown
to be effective over tens of kiloparsecs or for times of order $10^{8}$
years, and because specific models with specialized geometry are
needed to address any particular confinement picture,
this alternative will not be considered further here. 

\subsection{Inferences from Data}

The general relations derived above are applied to a subset of
radio bubbles and relic radio sources for which radio and x-ray
luminosities and geometries are known.  The source sample is taken
from the sample of objects discussed by Birzan et al. (2004).
These authors tabulate, for a sample of 17 sources,
radio and x-ray data, lobe geometries, $pdV$ values, buoyant
rise times and other derived properties for each component of
the sources in the sample.
The sample used here is taken from this single source in order to
ensure a uniform treatment in the derivation of radio lobe volumes,
buoyant risetimes and values of $pdV$. 
The objective is to consider 
clear examples of radio source configurations that have been 
inflated in the ICM, with well defined boundaries, 
to enable
the use of buoyant timescales in calculating energy fluxes and in
comparing various possible implications for the content of the
collimated outflows.  Hence the  non-cluster objects HCG 62 and M84
listed by Birzan et al. are not included here.  In 
addition the clearly "non-bubble" sources Cyg A
and M87 are also excluded.  Added to this list is the very
luminous radio bubble source MS0735 at a redshift of $0.22$ reported by
McNamara et al. (2005).  The final list of objects used is, in order
of decreasing luminosity, MS0735, Hydra A, A2597, MKW 3S, A2052,
A133, A4059, A2199, Perseus, RBS797, A1795, A478, 2A0335, and A262.
In each case the associated galaxy cluster name has been used,
consistent with the notation of Birzan et al. (2004).

For each of these sources, or for each component of a source
if the source is listed by Birzan et al. as having two components, 
the following quantities were calculated:
(1) The distance, using $H_{o} = 75$ km/sec/Mpc;
(2) the volume of each component, using the major and minor axis
data of Birzan et al. and assuming each component is an oblate
spheroid;
(3) the equipartition value of the magnetic field, using Eq. 2 for
equipartition with the electron population only;
(4) the energy in the electrons required to produce the radio
luminosity reported by Birzan et al., using Equations 1 and 2 
(for the source MS0735 the radio luminosity was calculated 
using a spectral index of $\alpha = 2.5$ (McNamara 2005)).
All calculations were carried out using a low frequency cutoff 
of 10 MHz (the same cutoff used by Birzan et al. (2004)) and also of 
3 MHz to test the sensitivity of the results to $\nu_{L}$;
(5) the radiative lifetime $t_{rad}$ against synchrotron losses at  
a frequency of $1$ GHz;
(6) the radio luminosity of each component as determined by the ratio
of its volume to the total volume times the total radio luminosity;
(7) the total energy of each component 
required to radiate at the observed luminosity
for a time $t_{rad}$ in order to compare with the $pdV$ energy;
(8) the total energy required for each component to radiate at
its luminosity for the buoyant risetime $t_{buoy}$ given by
Birzan et al.;
(9) the ratios of the energies (4), (7) and (8) to the $pdV$ energy of
that component given by Birzan et al.;
(10) the maximum total energy flux required for each component
to supply the $pdV$ energy in a time $t_{rad}$ and in a time 
$t_{buoy}$;
(11) the total energy density $\epsilon_{j}$ in a jet required by (10) for a 
given jet radius and bulk flow speed (from $dE_{j}/dt$)
for each component;
(12) the value of $nkT$ in the ambient ICM surrounding the jet.
This ambient value is found by use of a power law fit to
the radial dependence of ICM pressure and then determining the value
of $nkT$ in the ambient ICM at a distance one half that of the distance
from the nucleus of the parent galaxy to the center of the radio lobe.
This distance should approximate well the conditions that
surround the jet which feeds the radio lobe. 
The values of $nkT$ at the radio lobe are taken from the
results of Birzan et al. (2004).  The power law fit   
used is $p(r) = p_{0}(r/r_{0})^{-0.6}$, which fits well the data
reported by McNamara et al. (2005) for 
MS0735 as well as the data reported
for Perseus by Fabian et al.(2006).

\subsubsection{Total Energies}

The first comparison to make is one between the known $pdV$
energy required to inflate the radio lobes and the other estimates
of the energy present in each radio source component.
These other energy estimates are the 
calculated radio luminosity $L_{R}$
times the radiative lifetime using the equipartition magnetic
field discussed above, the radio luminosity times the estimated
buoyant risetime, and the minimum (equipartition) energy in
electrons required to produce the observed radio luminosity.  The
total minimum equipartition energy present is of course at least
twice this last value when the magnetic field is included, and
it may be more if the charge neutralizing species carry a significant
amount of energy.  But the issue here is a comparison of $pdV$
with estimates of the energy needed to produce the observed 
radio emission.  The $pdV$ energy is an observed minimum; there
must be at least this much energy carried out to each radio 
component.
\begin{figure}[h]
\figurenum{1}
\includegraphics[angle=-90,scale=0.3]{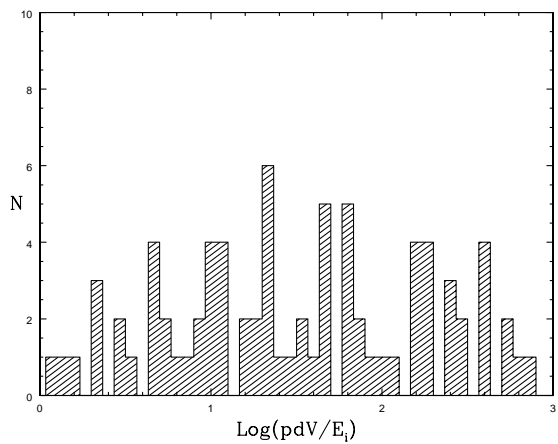}
\caption{Distribution of the ratio $(pdV/E_{i})$ for each component
of the radio sources in the sample.  Values of $pdV$ are taken from
Birzan et al. (2004), and $E_{i}$ are the values of $L_{R}t_{rad}$,
$L_{R}t_{buoy}$ and $E_{min}$ as described in the text.\label{fig1}}
\end{figure}

Figure 1 shows the distribution of the logarithm of $(pdV/E_{i})$,
where $E_{i}$ is each of the three other energies listed
above, for each radio component of the sample used here.
Hence each two component source contributes six entries to Fig. 1.
Two things are immediately evident from this figure.  First, in
{\it every} case the minimum total energy required to account for
the $pdV$ work exceeds all other energy estimates.
Second, the excess in $pdV$ energy over the other estimates is
often very large, from factors of ten to factors of several
hundred.\footnote{The values of $pdV/E$ for the powerful source
MS0735 lie near the upper end of the distribution in Figure 1,
and this is also true for the energy density distributions
discussed in Section 2.4.2 and Section 3. Although MS0735
is an exceptional source in this sample, it may be more
representative of FR-II objects and the most luminous FR-I
sources; hence it is of particular interest because of this.}
Thus this result is insensitive to uncertainties 
in the data introduced by measurement errors in the
lobe geometries or low frequency cutoffs in the synchrotron
spectrum, since these will introduce uncertainties of factors
of two or less.
The essential conclusion from this comparison is that
the energy present in the radio lobes required to account for the
$pdV$ work is by far the dominant energy measure for all these radio
sources, and possibly for all radio sources as well.  And again,
the $pdV$ energy is an {\it observed} quantity, not an estimate
based upon any assumptions about equipartition, source lifetimes,
or source dynamics.

\subsubsection{Jet Energetics}

What are the implications of this dominant $pdV$ energy for the
jets that supply this energy to the lobes?  Using the $pdV$
values, the required energy flux in the jet and an accompanying
energy density in the jet can be found, as described in steps
(10) and (11) above.  Calculation of an energy density requires
specification of an average jet cross sectional area ($\pi r_{j}^{2}$
) and a mean value of the speed $v_{j}$ with which material is
advected outward by the jet.  Nominal values chosen here are
$v_{j} = 0.1 c$ and $r_{j} = 0.1$ kpc.  The value of of $v_{j}$
was chosen because these objects are more similar to FR-I
objects than to FR-II sources, and hence the mean jet speeds are
likely to be non-relativistic or even subsonic or transonic.
In addition, in those cases where jet structures are seen in the
radio data, they are often distorted, which further suggests
non-relativistic speeds.  For completeness the calculations were
also carried out with $v_{j} = 0.5c$, but it seems that a mean
value of $v_{j}$ this large is unlikely.

The value of $r_{j}$ chosen is comparable to or larger than many
values of $r_{j}$ for highly resolved jets such as that in M87
(Sparks et al. 1996).  For most of the radio bubble sources in
this sample the resolution of the radio observations is not high
enough to resolve the jet structures that carry energy to the lobes.
Hence values for small scale jets such as M87 may be the most
appropriate here.  An extreme upper limit would be to use an average
jet diameter of 1 kpc, which is applicable as an upper limit to
very large scale ($\sim 100$ kpc) jets such as Cygnus A (e.g.,
Swain, Bridle \& Baum 1996).  Values of jet energy density 
$\epsilon_{j}$ for this limiting value of jet diameter were also
calculated.  Hence the nominal values of $v_{j} = 0.1c$ 
and $r_{j} = 0.1kpc$ are near the middle to 
upper end of the range of values applicable to 
these objects (with the possible exception of MS0735), and
this choice will then result in values of the jet energy density
that are near the lower end of their possible range.

The resulting values for $\epsilon_{j}$ can then be compared to the
values of $nkT$ for the ambient ICM surrounding the jets, as 
described in steps (11) and (12) in the previous section.
Figure 2 shows the 
distribution of $\epsilon_{j}$ derived from the $pdV$ requirements
divided by the values of $nkT$ in the ambient ICM around each jet.
For $r_{j}=0.1 kpc$ and $v_{j} = 0.1c$, Figure 2 
shows that almost all (96\%) jets with an internal energy
density required to supply the observed $pdV$ energy to their
associated lobe cannot be confined by the external medium.  This
result is independent of the form taken by the energy being
transported by the jet and only requires that the 
$\epsilon_{j}$ produce pressure via
a standard equation of state.  
\begin{figure}[h]
\figurenum{2}
\includegraphics[angle=-90,scale=0.3]{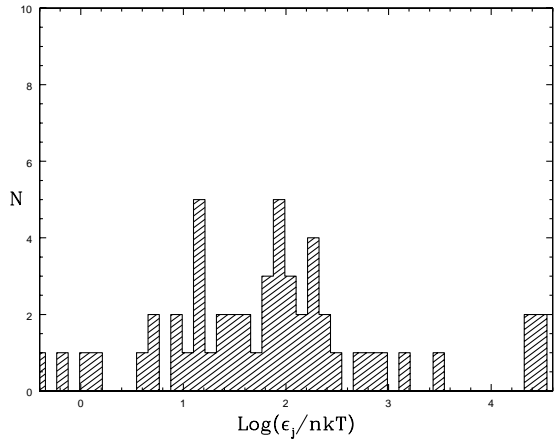}
\caption{Distribution of $(\epsilon_{j}/nkT)$ for each component
of the sample, where $\epsilon_{j}$ is the jet internal energy
from $pdV$ requirements and $nkT$ is the ambient pressure of
the ICM surrounding the jet. Here $v_{j}=0.1c$ and $r_{j}=0.1kpc$.
\label{fig2}}
\end{figure}

Thus, unless some specific
external magnetic confinement model is employed in almost every
cluster radio source, the jet energy densities required would
result in decollimation and destruction of the jet.  The resulting
outflow would be more like an isotropic wind and would not produce
the typical double lobed radio structure seen here and in essentially
all extended radio sources.
Increasing the mean jet advection speed to $0.5$c for all jets
decreases the value of $\epsilon_{j}$, but in most cases the jet
energy densities still cannot be confined.
In this case 90\% of the jets could not exist if they were to supply
the required energy to their respective lobes.  Increasing the jet
diameter to 1 kpc lowers the internal energy more dramatically,
but even in this case almost two-thirds (65\%) of the jets
would not be confined.  It is important
to recall that the $pdV$ values are minimum energy requirements
because the enthalpy in the lobes is larger than this, and in
addition $t_{rad}$ and $t_{buoy}$ are upper limits to 
the actual outflow times. 

\section{Implications for Jet Content}

The results of Section 2.4.2 show that measurements of the $pdV$
work needed to create the radio "bubble" sources in the sample used here 
imply that the jets suplying the energy will be decollimated in almost
all cases as long as the energy density in the jet has an associated
pressure given by a conventional equation of state.
One obvious solution to this problem is to
include in the content of the jet a component that carries energy
but does not contribute significantly to the internal pressure of
the jet.  A leading candidate for this component is a population
of "cold" protons as the charge neutralizing species.  

A cold proton population moving at the advection speed of the jet
can carry energy as directed kinetic energy $m_{p}v_{j}^{2}$ per
particle, and this energy can be significantly greater than the 
energy per particle carried by the relativistic electrons.  If
the relativistic electron population has a power law energy
distribution of the form $N(E)dE = K E^{-p}dE = K_{1}\gamma^{-p}
d\gamma$, where $\gamma$ is the Lorentz factor of the electrons,
then the average value of $\gamma$ is just
\begin{equation}
{\bar \gamma} = \frac{\int N(\gamma)\gamma d\gamma}{\int N(\gamma)d\gamma} =
\frac{\int \gamma^{1-p} d\gamma}{\int \gamma^{-p} d\gamma}.
\end{equation}
For $p > 2$,
${\bar \gamma} = \frac{(p-1)}{(p-2)} \gamma_{min}$ 
(using 
$\gamma_{max} >> \gamma_{min}$),
and for a radio flux of the form $S_{\nu} \propto \nu^{-\alpha}$,
$p = 2 \alpha + 1$, hence values of $p \ge 2$, corresponding to
$\alpha \ge 0.5$, are most likely here.  The value of $\gamma_{min}$
is not known, though estimates can be obtained via inverse Compton
scattering models of x-ray emitting jets (e.g., Celotti et al. 2001).
Conservative estimates of $\gamma_{min}$ in the range of 1 - 10,
together with $\gamma_{max}$ values of $10^{3-4}$, all suggest
that ${\bar \gamma} \sim 10$ is a number with wide applicability.  In any
case $\bar \gamma$ is very unlikely to be near $10^{3}$, and hence
the cold protons carry much more energy in the jet than the 
electron population.  The interaction of these protons with the
relativistic electrons in the jet will be considered in Section 4,

For the sample of sources here one can equate the required energy
in $pdV$ work to an energy flux 
carried by cold protons in the jet, ${\dot E}_{p} = 
\pi n_{p} m_{p} v_{j}^{3} r{j}^{2}$, times the outflow lifetime.
This then gives the number
density of cold protons in the jet needed to provide the
required energy.  The jet proton number densities required
for each of the radio sources used here fall between $0.2$ and
$3 \times 10^{-6}$, with a concentration between $0.02$ and
$x \times 10^{-4}$ cm$^{-3}$, using 
energy fluxes delivered over the electron radiative lifetimes or
over the buoyant risetimes and with 
$v_{j} = 0.1c$ and $r_{j}=0.1kpc$. 
For the much higher average
jet speed of $v_{j}=0.5c$, the values of $n_{p}$ will be reduced
by a factor of 125.  
The directed flux of protons can be thermalized
through one or more termination shocks that occur at the end of the
jet when it encounters either the surrounding ICM and begins to form
the radio lobes or when it encounters a similar shock due to 
the internal back pressure in the already growing lobe. 

The simplest cold proton flux model assumes that the
relativistic electrons are also advected outward at $v_{j}$
along with a dynamically unimportant magnetic field and that the
electron momentum distribution is isotropic in a co-moving frame.
Setting the
electron number density $n_{e}$ equal to the derived proton
number density $n_{p}$ and assigning an average $\bar \gamma$
to each electron then allows calculation of the electron energy
density $\epsilon_{e}$ within the jet and also the total electron
energy flux ${\dot E}_e$ into the lobe.  A first test is to see
if the values of ${\dot E}_e$ are equal to or greater than the
radio luminosity $L_{R}$ associated with a given lobe.  The calculated
values of ${\dot E}_{e}/L_{R}$ here 
for $v_{j} = 0.1c$ and $r_{j} = 0.1$ kpc
are greater than one for all but 3 out of 52 source components,
so in general this requirement is satisfied.
A second test is to compare values of $\epsilon_{e}$ with the
values of $nkT$ in the ICM surrounding the jets.  With an average
value of ${\bar \gamma} = 10$, jet parameters of $v_{j}=0.1c$ and
$r_{j} = 0.1$ kpc give values of 
$\epsilon_{e} = n_{e}{\bar \gamma} m_{e} c^{2}$ that are greater
than the ambient values of $nkT$ for almost all radio components
(48/52), resulting in jet destruction in those cases.  
The distribution of $\epsilon_{e}/nkT$ 
in this case is shown in Figure 3. 
(In principle there is a contradiction between the value of
$\bar\gamma = 10$ and a value of $\nu_{L} \sim 10^{6}$ Hz
used earlier, but only if $\nu_{L}$ reflects a true cutoff in
the radio emission, which is unlikely.  Raising $\bar \gamma$
to correspond to $\nu_{L}$ means $\bar \gamma \sim 10^{2}$,
which makes the confinement problem even more severe.)
If the average jet speed is increased to $v_{j} = 0.5c$, the required
values of $n_{p} = n_{e}$ decrease by a factor of 125, as does
$\epsilon_{e} = n_{e}{\bar \gamma} m_{e} c^{2}$.  In this case
jet confinement and collimation can be maintained for about 80\%
of the objects.  Increasing the jet
diameter to 1 kpc will ensure confinement in almost all cases.
\begin{figure}[h]
\figurenum{3}
\includegraphics[angle=-90,scale=0.3]{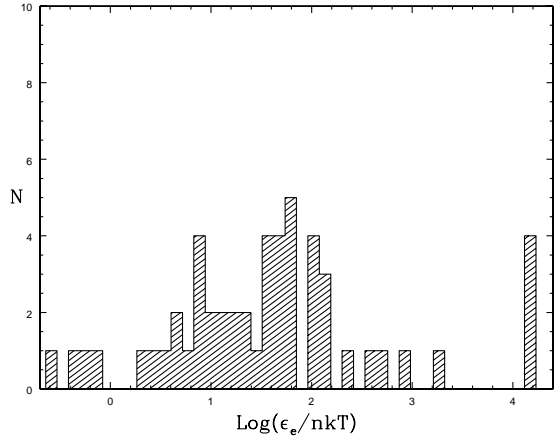}
\caption{Distribution of the ratio of electron energy density
$\epsilon_{e}$ (with $n_{e}=n_{p}$) in each jet to the ambient
ICM pressure $nkT$.  In this case $\bar\gamma=10$, $v_{j}=0.1c$,
and $r_{j}=0.1kpc$.\label{fig3}} 
\end{figure}

However, high speed flow all along 
the jet creates a further complication
if the electrons flowing in the jet are the sole supply of 
energy that powers the observed synchrotron radiation.  For mean
values of $v_{j} = 0.1c$, the energy flux in electrons emerging from
the jet is comparable to or greater than the integrated radio
luminosity of the associated lobe in nearly every case, as
described above.  Since the net proton energy flux ${\dot E}_{p} =
\pi r_{j}^{2} n_{e} m_{p} v_{j}^{3}$ is fixed by the $pdV$
requirement in a given lobe, and since $n_{e} = n_{p}$, 
increasing the flow speed to $0.5c$
to avoid decollimation reduces the jet electron energy density
$\epsilon_{e} = n_{e} {\bar \gamma} m_{e} c^{2}$ at a fixed 
$\bar \gamma$ by $(\Delta v_{j})^{3}$,
while the flux of particles from the jet into the lobe increases
only by $\Delta v_{j}$.  As a result, the net electron energy flux 
${\dot E}_{e} = \pi r_{j}^{2} \epsilon_{e}  v_{j}$ into 
the lobe is reduced, and in the case of $v_{j}= 0.5c$, the electron
energy flux ${\dot E}_{e}$ is equal to or greater than the radio
luminosity of the associated lobe in only 46\% of the lobes in this
sample.  Thus high values of $v_{j}$ may require some mechanism
to augment the energy of the relativistic electron population,
possibly via conversion of some fraction of the proton energy flux.

Hence an important conclusion about cold proton dominated jets
is that such jets can provide the very large energy fluxes needed
to explain the values of $pdV$ energy observed in the ICM cavities,
but if the accompanying relativistic power law distribution of
electron energy is isotropic in the frame comoving with the jet, then
unrealistically large values of the jet average flow speeds are
needed to avoid jet decollimation.  In addition, such high flow
speeds in turn reduce the net electron energy flux into the
lobes to a level that cannot support the observed radio luminosities
in roughly two-thirds of the sources in the sample.  These results
may indicate the need for more complex internal structures in
particle dominated jet models beyond the simple structures tested
here, such as conversion of
some fraction of the proton energy to the relativistic electron
population or use of a "cold but relativistic" electron beam.
It is not clear how this latter option could be maintained 
against streaming instabilities or
how it would produce the synchrotron radiation seen in jets.
The problems are
generally most extreme for large and luminous sources such as 
the one seen in MS0735 (and possibly all FR-II objects as well).

A further indication of the possible need for complexity in
particle dominated jet flows comes 
from a minimalist electron energy
flow model that is independent of all the above restrictions 
imposed by $pdV$ energy requirements.  Without reference to 
total energy fluxes, possible proton populations, etc., an
electron energy flux required to produce the observed radio
emission alone can be constructed from the ratio of the 
electron energy $E_{e}$ divided by the
radiative lifetime $t_{rad}$ of the same electrons
(cf. Sect 2.3).
This ratio is independent of the magnetic field strength
and is thus not dependent on any assumptions about equipartition
magnetic fields.  From Eqs. 1 and 3,
\begin{eqnarray}
\dot E_{e,min} = (E_{e}/t_{rad}) = 
  \frac{8 \pi}{27 a(p)}D^{2} \nu^{1/2} \times \nonumber \\
\frac{S_{o}\nu_{o}^{\alpha}}
{1/2 - \alpha}(\nu_{U}^{(1/2-\alpha)}
- \nu_{L}^{(1/2 - \alpha)}).
\end{eqnarray}

This ratio gives a minimum value for the electron energy
density $\epsilon_{e,min}$ in the jet for a given value of $v_{j}$
and $r_{j}$.  Values of $\epsilon_{e,min}$ can be found for each
radio component and compared again to the ambient pressure in the
ICM that surrounds the jet.  The distribution of these ratios 
shows that the internal energy
density from this minimum electron energy flow at $v_{j}=0.1c$ 
and $r_{j}=0.1$ kpc still
exceeds the ambient pressure for many objects (58\%). 
Increasing the flow speed to $0.5c$ and 
jet diameter to 1 kpc 
essentially removes this problem, 
but only for extremum values of jet flow
speed and size, given the overall geometry of these small radio
sources.  This exercise again suggests that a simple 
particle dominated jet picture may encounter difficulties even in
the absence of the increased energy requirements imposed by the
$pdV$ measurements.  In principle a "beamed" population of very cold  
relativistic particles, in addition to a heavy cold particle flux,
could, if stable, satisfy all the constraints on energy supply and 
collimation.  A beamed electron-positron jet would be
less successful in solving this problem unless the average $\gamma$
of the electrons (and positrons) is of order 1000, which in turn
places strong constraints on either the spectral index $\alpha$ or
the value of the low energy cutoff $\gamma_{min}$ of the particle
energy distribution.

\section{Internal Dynamics of Electron-Proton Jets}

A question that has not generally been addressed in particle dominated
jet models is that of the interactions among particle species present
in the jet.  Of particular relevance here are the particle interactions 
in a "cold proton - hot electron" jet that can provide the required
$pdV$ energy and the radio luminosity as described in Section 3.  The
basic question is whether or not the relativistic electrons can
transfer a significant amount of their energy to 
the comoving cold protons during their time of residence
in the jet.  This is an issue of general interest in jets with 
multiple particle populations at different temperatures; in this case
if there is a significant energy transfer then not only may the 
electrons be unable to power the observed radio luminosity without
reacceleration, but there may in addition be important changes in
the power law form of the electron energy distribution.  Also, if
the cold proton population is "warmed" in any significant manner
then the interactions among protons, electrons and magnetic field
become more complex, and the possibility of significant thermalization
of the proton population may arise, resulting in potential 
decollimation and destruction of the jet.

Thus the problem is to calculate the collision rate and resulting
energy equipartition time between relativistic electrons and cold
protons.  A similar problem involving the interaction between very
high temperature "thermal" electrons and cooler protons has been
treated by Gould (1981, 1982).  The problem here is somewhat different
in that it involves the interaction between a power law distribution 
of electron energies.  However, the approach here uses the method
developed by Gould as a starting point.  The first step is to calculate
the energy loss of a relativistic electron population against
electron-proton collisions; from this an energy loss 
time $[(1/E)dE/dt]^{-1}$ 
can be found and compared to the estimated lifetimes of the radio
source obtained from $t_{rad}$ or $t_{buoy}$. (The treatment in
this section neglects the presence of a magnetic field.  Though
electron gyroradii are small compared to the radio source
dimensions, electron trajectories at high energies can still
be treated as single particle paths moving in a stationary
background of cold protons.)

A good approximation in the co-moving jet frame is to consider the
target protons as stationary, since their temperature is very low.
In this case the energy loss rate of an incident relativistic
electron is just $\Delta E_{e}/\Delta t 
= n_{p} \int \Delta E_{p} v_{e} d\sigma$,
where $\Delta E_{p}$ is the energy transferred to the proton, $v_{e}$
is the electron speed, and $d\sigma$ is the differential scattering
cross section.  This interaction is basically that of relativistic
Coulomb scattering with quantum mechanical corrections, and 
integration of the
above equation for $\Delta E_{e}/\Delta t$ 
gives, in the Born approximation (Gould 1981),
\begin{equation}
\frac{\Delta E_{e}}{\Delta t} = - {{4 \pi e^{4}}\over m_{p} c} n_{p} 
[ln(2 E_{e}/\hbar \omega_{p}) - 1/2].
\end{equation}
The above equation has set $\beta_{e} \simeq 1$, and the $\omega_{p}$
term arises from very small angle, low energy transfer scattering 
that can be represented by a single plasma oscillation.  Since here
the relativistic electrons have a non-thermal energy distribution,
$\omega_{p}^{2} \simeq 4 \pi n_{e}e^{2}/{\bar \gamma} m_{e}$,
where ${\bar \gamma}m_{e} \ll m_{p}$, $n_{e}$ is averaged over
electron energy, and $\bar \gamma$ is the average value of the
Lorentz factor for a power law energy distribution.  Charge neutrality
in the jet requires $n_{e} = n_{p}$, and Eq. 8 then allows an 
evaluation of the electron energy loss time, 
$t_{e-p} = [1/E_{e}(dE_{e}/dt)]^{-1}$.
Evaluating the coefficients gives
\begin{equation}
t_{e-p} = {{1.9 \times 10^{9} \gamma_{e}/n_{p}}\over {37.5 +
ln[\gamma_{e}/(n_{e}/{\bar \gamma})^{1/2}]}},
\end{equation}
where the units are years.

Average number densities in the proton dominated jets here range from
about 0.01 to $10^{-6}$, with an average near $10^{-3}$. 
Hence it is clear 
from Equation 7 that the high energy electrons responsible for the
radio emission ($\gamma_{e} > 10^{3}$) do not lose any
significant energy to the proton population over the source lifetime.
For spectral indices of $\alpha = 0.5$ or greater, most of the
electrons have values of $\gamma_{e}$ near $\gamma_{min}$, but even
with $\gamma_{e} \sim 10$ and $n_{p} \sim 0.1$ the electron-proton
loss times are well in excess of the upper end of the distributions 
of source lifetimes, which lie at values of $t_{buoy} \sim 10^{8}$ yr.
Thus in general there is a negligible transfer of energy within the
jet from a relativistic electron population with a power law energy
distribution to a cold proton population.  Hence the "cold proton - 
hot electron" model for particle dominated jets can maintain the two
distinct populations over the transit time for particles in the jet.

Though not directly related to cold proton dominated jets, another
possible energy transfer process that has not been treated previously
in extragalactic radio jets is the loss of energy 
due to electron-electron collisions from the high
energy end of a power law electron energy distribution 
to the much more numerous population of low energy 
electrons near the low energy end of the distribution.  If this 
interaction is significant, it could have a very large effect on
the evolution of the radio luminosity of extended radio sources.
Again following Gould (1981), for an isotropic distribution
of electrons the loss rate of particle $1$ against a lower energy
population of particles of energy $E_{2}$, where 
$n(E_{2}) = K E_{2}^{-p}$, gives upon integration 
\begin{equation}
-\frac{1}{\gamma_{1}} \frac{d\gamma_{1}}{dt} =
2 \pi r_{o}^{2}c n_{e} \frac{p-1}{p} [ln(\gamma_{1}/\gamma_{L}) -
1/p - 0.82],
\end{equation}
where $\gamma_{1}$ is the Lorentz factor of the high energy electron,
$\gamma_{L}$ is the low energy end of the electron energy
distribution, and $r_{o}$ is again the classical electron radius.
The condition $\gamma_{H} \gg \gamma_{L}$ has been
used in deriving Eq. 8.  Values of the electron energy transfer
time $t_{e-e} = [1/\gamma_{1}(d\gamma_{1}/dt]^{-1}$ can then be
found.  Evaluating the coefficients in Eq. 10 then gives
$$
t_{e-e} = 6.76 \times 10^{13} n_{e}^{-1} \frac{p}{p-1}
[ln(\gamma_{1}/\gamma_{L}) -
1/p - 0.82]^{-1}
$$
in cgs units.  For $p \simeq 3$, $\gamma_{1} \ge 10^{5}$,
and $\gamma_{L} \simeq 10$, this reduces to 
\begin{equation}
t_{e-e} \simeq 2 \times 10^{6} n_{e}^{-1} yr.
\end{equation}
For the proton dominated jets with $n_{e} = n_{p}$, the values
of $n_{e}$ in the present sample are less than $10^{-2}$ in
almost all (85\%) of the cases for $v_{j}=0.1c$
and $r_{j}=0.1kpc$, and the maximum value of $n_{e}$ in this
case is $1.45 \times 10^{-2}$.  For larger values of $v_{j}$
and $r_{j}$, $n_{e}$ will be much smaller.
Thus it is clear that for almost all values derived
for $n_{e}$, the transfer of energy from the high energy electrons
to the more numerous low energy population is negligible over
the estimated lifetimes of the radio sources.

\section{Summary and Conclusions}

\subsection{Summary}

Examination of data from recent radio and x-ray observations of
radio sources in clusters of galaxies shows that the energy
required to inflate the radio lobes against the ambient pressure
of the ICM exceeds all other (and less certain) estimates of the
energy present in the radio lobes. 
Supply of this $pdV$ energy to the radio lobes
places a stringent {\it lower limit} on the energy flux that
must be supplied by the jet over any given time, since the enthalpy
of the material residing in the lobe exceeds this $pdV$ value.
Specification of a jet radius $r_{j}$ and a mean flow speed $v_{j}$
allows a determination of the energy density (in an unspecified
form) that must be present in the jet, and it was shown in Section 3
that this energy density $\epsilon_{j}$ almost always exceeds
the known ambient pressure of the ICM surrounding the jet.  Hence
such jets cannot be pressure confined, and either some other
confining mechanism such as magnetic confinement must be employed,
or else the energy density in the jets must not have a significant
associated pressure.
We conclude that 
the most natural way to do this is to use a charge neutralizing
population of "cold" protons that are advected outward with the
mean jet flow.  For a given jet radius and mean jet flow speed,
the energy flux in protons required to provide the $pdV$ energy
over either a radiative or a buoyant lifetime then fixes the 
particle number density ($n_{p} =n_{e}$) in the jet.  This then
specifies the mean jet kinematics.

Although cold proton dominated jets can solve the required energy
flux problem, the accompanying relativistic electron population
can still pose difficulties.  The derived proton number density $n_{p}$
gives the electron density $n_{e}$, and the jet parameters plus
the radio source luminosity and radiative or buoyant lifetimes
then give the electron energy density $\epsilon_{e}$ in the jet.
The results of Section 3 show that in many cases this lower value of 
$\epsilon$ ($\epsilon_{e} < \epsilon_{p}$) also exceeds the values of
the ambient pressure unless large (and possibly unrealistic)
values of $v_{j}$ and $r_{j}$ are used, and thus jet destruction
will occur.  Section 3 also showed that even the minimum 
energy present in the jet can cause jet destruction in a
significant number of cases.  Another conclusion from Section 3
is that electron-positron jets will have great difficulty in 
meeting the overall energy transport requirements unless extremely
cold and extremely relativistic ($\bar \gamma
\ge 1000$) energy distributions are used.

In Section 4 the interaction among particle species in particle
dominated jets was calculated, and it was shown that in almost
all cases the cold proton-relativistic electron energy exchanges
are so small that negligible energy transfer between these two
populations will occur over the source lifetime.  In addition,
energy transfer between the high and low energy ends of a power
law electron energy distribution present in a jet was calculated.
In this case it was also found that this energy transfer has no
significant effect over the lifetimes of the radio sources
considered here.

\subsection{Conclusions}

Cold proton particle dominated jets can transport the required
energy to the radio lobes.  But the results of Section 3 suggest
that the accompanying electron energy 
density in radio jets may cause decollimation 
unless some form of magnetic confinement is used.
In the absence of this, some version of electron streaming that
reduces the transverse pressure 
may be needed, but it is not clear how to 
construct such distributions in a natural way in the first place,
nor is it clear how then to produce the synchrotron radiation
seen in jets or how 
to protect this streaming population against
instabilities with rapid growth rates.

A major
unresolved issue is that of the creation and acceleration of the
cold proton population.  Entrainment of a (relatively) cold ISM
through the mixing layer on the surface of the jet may work,
especially since Section 4 showed there is little interaction
between the hot electrons and cold protons.  However, this 
entrained ISM is charge neutral, and so the charge neutralizing
species in the original jet has still to be specified.  Moreover,
the initial jet must still carry all the energy (in some form)
needed to supply the $pdV$ energy in the lobes; transfer of this
energy to a cold entrained proton population only adds the need
for an additional process to take place in the jet.  Electron-positron
jets cannot solve this problem unless they are 
cold and highly beamed, 
as discussed in Section 3.  Thus once again the solution to the
problem is pushed back to the largely unknown physical processes
occurring in the immediate neighborhood of the central black hole.
All of these problems may be more serious for the general population
of FR-II objects; they have very large lobe energies and thus
large $pdV$ values, and they usually have very highly collimated
jets that extend over very long distances.

An alternate solution that may address these issues is the class of
"Poynting Flux" models of collimated
energy transport which have a nearly negligible particle content
and are dominated by an intense and highly collimated
electromagnetic flux 
(e.g., Romanova et al. 1997; Ustyugova et al. 2000;
Lovelace et al. 2002; Lovelace \& Romanova 2003).
These models are still under development, and it is not clear
at this time if they can provide stable, long term
collimated outflows over long distances that can reproduce the
radio morphology of both lobes and jets.  In particular, major
difficulties may be encountered in reproducing the morphologies
of the most common FR-I class of objects.  However, the results 
shown here may serve to motivate additional development of these
models as well as encourage possible refinement and modification
of particle dominated jet models.

\acknowledgments

Thanks are due to Tom Jones and Brian McNamara for valuable comments
and to a referee for very useful suggestions.





\clearpage

\end{document}